\documentstyle[ltwol,epsbox,graphicx,here]{article}

\def\Journal#1#2#3#4{{#1} {\bf #2}, #3 (#4)}

\def\PLB{{\em Phys. Lett.}  B}
\def\PRL{\em Phys. Rev. Lett.}

\def\ZPC{{\em Z. Phys.} C}

\def\be{\begin{equation}}
\def\ee{\end{equation}}
\def\bea{\begin{eqnarray}}
\def\eea{\end{eqnarray}}

\def\tdzresultns{414.5\pm1.7(stat.)}
\def\tdpresultns{1029\pm12(stat.)}
\def\tdsresultns{488.4^{+7.8}_{-7.7}(stat.)}
\def\rdpresultns{2.48\pm 0.03(stat.)}
\def\rdsresultns{1.18\pm 0.02(stat.)}
\def\yresultns{1.16^{+1.67}_{-1.65}(stat.)}
\def\intl{11.1}

\def\tdz{\tau(\dz)}
\def\tdp{\tau(\dplus)}
\def\tds{\tau(\ds)}
\def\rdp{\tdp/\tdz}
\def\rds{\tds/\tdz}

\def\Tdzresultns{$\tdz=(\tdzresultns)$ fs}
\def\Tdpresultns{$\tdp=(\tdpresultns)$ fs}
\def\Tdsresultns{$\tds=(\tdsresultns)$ fs}
\def\Rdpresultns{$\rdp=\rdpresultns$}
\def\Rdsresultns{$\rds=\rdsresultns$}
\def\Yresultns{$\ycp=(\yresultns)$ \%}
\def\etal{{\it et al.}}
\def\ycp{y_{CP}}
\def\am{A_{mix}}

\def\Fb{fb$^{-1}$}

\def\qqbar{q\overline{q}}
\def\ccbar{c\overline{c}}
\def\uubar{u\overline{u}}
\def\ddbar{d\overline{d}}
\def\ssbar{s\overline{s}}

\def\pip{\pi^+}
\def\pim{\pi^-}
\def\piz{\pi^0}
\def\km{K^-}
\def\kp{K^+}
\def\kstarzb{\overline{K}^{*0}}

\def\dstarp{D^{*+}}

\def\dz{D^0}
\def\dplus{D^+}
\def\ds{D_s^+}
\def\Dz{$\dz$}
\def\Dp{$\dplus$}
\def\Ds{$\ds$}

\def\kpkm{\kp\km}
\def\kmkp{\km\kp}
\def\kpi{\km\pip}

\def\kpipi{\km\pip\pip}

\def\dzkpi{\dz\to\kpi}
\def\dzkk{\dz\to\kmkp}
\def\dpphipi{\dplus\to\phi\pip}
\def\dpkpipi{\dplus\to\kpipi}
\def\dsphipi{\ds\to\phi\pip}
\def\dskstk{\ds\to\kstarzb\kp}

\def\UPS{$\Upsilon(4S)$}
\def\Gevc{GeV/$c$}

\def\Mevcsq{MeV/$c^2$}

\def\tsig{\tau_{SIG}}

\def\ie{{\it i.e.}}
\def\pt{p_T}

\def\vec2#1{\mbox{\boldmath $#1$}}

\begin{document}

\title{MEASUREMENTS OF CHARMED MESON LIFETIMES AND
SEARCH FOR $D^0$-$\overline{D}^0$ MIXING WITH THE BELLE EXPERIMENT}

\author{J. TANAKA}

\address{Department of Physics, University of Tokyo,
7-3-1 Hongo\\E-mail: jtanaka@hep.phys.s.u-tokyo.ac.jp}


\twocolumn[\maketitle\abstracts{The lifetimes of charmed mesons have been measured
using \intl~\Fb\ of data collected with the Belle detector at KEKB.
Each candidate is fully reconstructed to identify the flavor of the charmed meson.
The lifetimes are measured to be \Tdzresultns, \Tdpresultns\ and \Tdsresultns,
where the error is statistical only.
The ratios of the lifetimes of \Dp\ and \Ds\ with respect to \Dz\ are
measured to be \Rdpresultns\ and \Rdsresultns. 
The mixing parameter $\ycp$ is also measured to be \Yresultns\
through the lifetime difference of \Dz\ mesons decaying into
CP-mixed states and CP eigenstates.
All results are preliminary.}]

\section{Introduction}
Measurements of individual charmed meson lifetimes provide
useful information for the theoretical understanding of
the heavy flavor decay mechanisms\cite{life-th,life-ratio-th}.
Moreover, the $D^0$-$\overline{D}^0$ mixing parameters,
$y\equiv(\Gamma_H-\Gamma_L)/{(\Gamma_H+\Gamma_L)}$ and
$x\equiv 2(M_H-M_L)/(\Gamma_H+\Gamma_L)$, can be explored by 
measuring the lifetime difference of the \Dz\ meson decaying
into a CP-mixed state $\dzkpi$ and a CP-eigenstate $\dzkk$.
The parameter $\ycp$, defined by
$\ycp\equiv\frac{\Gamma(\mathrm{CP\ even})-\Gamma(\mathrm{CP\ odd})}
{\Gamma(\mathrm{CP\ even})+\Gamma(\mathrm{CP\ odd})}=\frac{\tau(\dzkpi)}{\tau(\dzkk)}-1,$
is related to $y$ and $x$ by the expression
$\ycp=y\cos\phi-\frac{\am}{2} x\sin\phi,$
where $\phi$ is a CP violating weak phase due to the
interference of decays with and without mixing, and $\am$ is
related to CP violation in mixing.
E791\cite{E791a}$^,$\cite{E791b}, FOCUS\cite{FOCUS} and CLEO\cite{CLEOb} have measured
$\ycp=(0.8\pm2.9\pm1.0)$\%, $\ycp=(3.42\pm1.39\pm0.74)$\% and
$\ycp=(-1.1\pm2.5\pm1.4)$\% respectively.
It is interesting that the FOCUS result is non-zero by more
than two standard deviations.
On the other hand, CLEO\cite{CLEOa} gives results for $D^0$-$\overline{D}^0$ mixing
through $\dz\to\kp\pim$, $y'\cos\phi=(-2.5^{+1.4}_{-1.6})$\%,
$x'=(0.0\pm1.5\pm0.2)$\% and $\am=0.23^{+0.63}_{-0.80}$, where
$y'=y\cos\delta - x\sin\delta$ and $x'=x\cos\delta + y\sin\delta$;
the parameter $\delta$ is the strong phase between the doubly
Cabibbo suppressed decay $D^0 \to K^+ \pi^-$ and the Cabibbo
allowed decay $\overline{D}^0 \to K^+ \pi^-$ ($\delta=0$ in the $SU(3)$ limit). 
The FOCUS and CLEO results could be consistent if there is
a large $SU(3)$ breaking effect in $\dz\to K^\pm\pi^\mp$ decays\cite{mix-th}.

\section{B-Factory}
\subsection{Accelerator: KEKB}
The KEKB\cite{KEKB} is an asymmetric energy
electron-positron collider designed to boost $B$ mesons.
The electron and positron beam energies are 8 GeV and 3.5 GeV
respectively: the resulting CMS energy, 10.58 GeV,
corresponds to the the mass of the $\Upsilon$(4S) resonance.
At the design luminosity, $10^{34}$cm$^{-2}$s$^{-1}$, about
$10^8$ $\Upsilon$(4S) are produced per year. So-called
continuum events($e^+e^-\to\uubar,\ddbar,\ssbar,\ccbar$) are also
produced, in the ratio $\qqbar$:$\Upsilon$(4S) $\simeq$ 3.5:1.

\subsection{Detector: Belle}
The Belle detector\cite{Belle} consists of
a three-layer silicon vertex detector(SVD),
a 50-layer central drift chamber(CDC),
an array of 1188 aerogel \v{C}erenkov counters(ACC),
128 time-of-flight(TOF) scintillation counters,
an electromagnetic calorimeter containing 8736 CsI(T$l$) crystals(ECL)
and 14 layers of 4.7-cm-thick iron plates interleaved with a system of resistive plate counters(KLM).
All subdetectors, apart from the KLM, are located inside
a 3.4-m-diameter superconducting solenoid that provides a
1.5 Tesla magnetic field.
The transverse momentum resolution for charged tracks is
$(\sigma_{\pt}/\pt)^2=(0.00019\pt)^2+(0.0034)^2$, where $\pt$ is in \Gevc\ and
the impact parameter resolution for $p$=1\Gevc\ tracks at normal incidence is
$\sigma_{r\phi}\simeq\sigma_z=55\mu$m.
$dE/dz$ measurements in the CDC($\sigma_{dE/dx}$=6.9\%),
TOF flight-time measurements($\sigma_{TOF}$=95ps) and
the response of the ACC provide $K^\pm$ identification with an efficiency
of about 85\% and a charged pion fake rate of about 10\% for all momenta up to 3.5 \Gevc.
Photons are identified as ECL showers that have a minimum energy of 20 MeV and are not
matched to a charged track. The photon energy resolution is
$(\sigma_E/E)^2=(0.013)^2+(0.0007/E)^2+(0.008/E^{1/4})^2$, where $E$ is in GeV.

\section{Reconstruction}
\Dz, \Dp\ and \Ds\ mesons are fully reconstructed
via the decay chains\footnote{Charge-conjugate modes are implied throughout this paper.}
$\dz\to\kpi$,
$\dz\to\kmkp$,
$\dplus\to\kpipi\ ({\mathrm with}\ \dstarp\to\dplus\piz\ {\mathrm requirement})$,
$\dplus\to\phi\pip$, $\phi\to\kpkm$,
$\ds\to\phi\pip$, and
$\ds\to\kstarzb\kp$, $\kstarzb\to\kpi$.

\begin{figure}[H]
(a)\hspace{0.1mm}\vspace{-6.5mm}
\begin{flushright}
\resizebox{0.45\textwidth}{0.25\textwidth}{\includegraphics{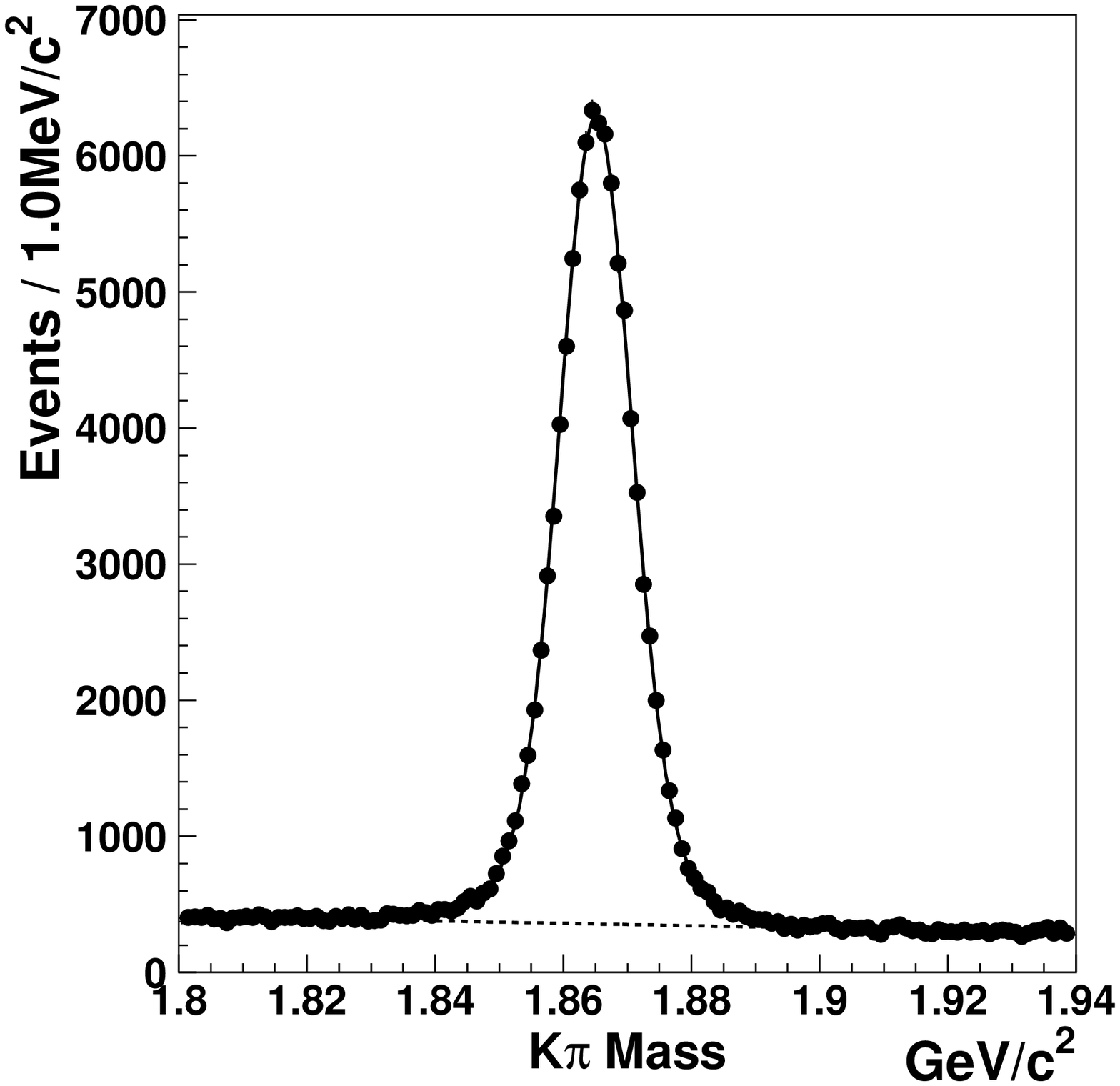}}
\end{flushright}
(b)\hspace{0.1mm}\vspace{-6.5mm}
\begin{flushright}
\resizebox{0.45\textwidth}{0.25\textwidth}{\includegraphics{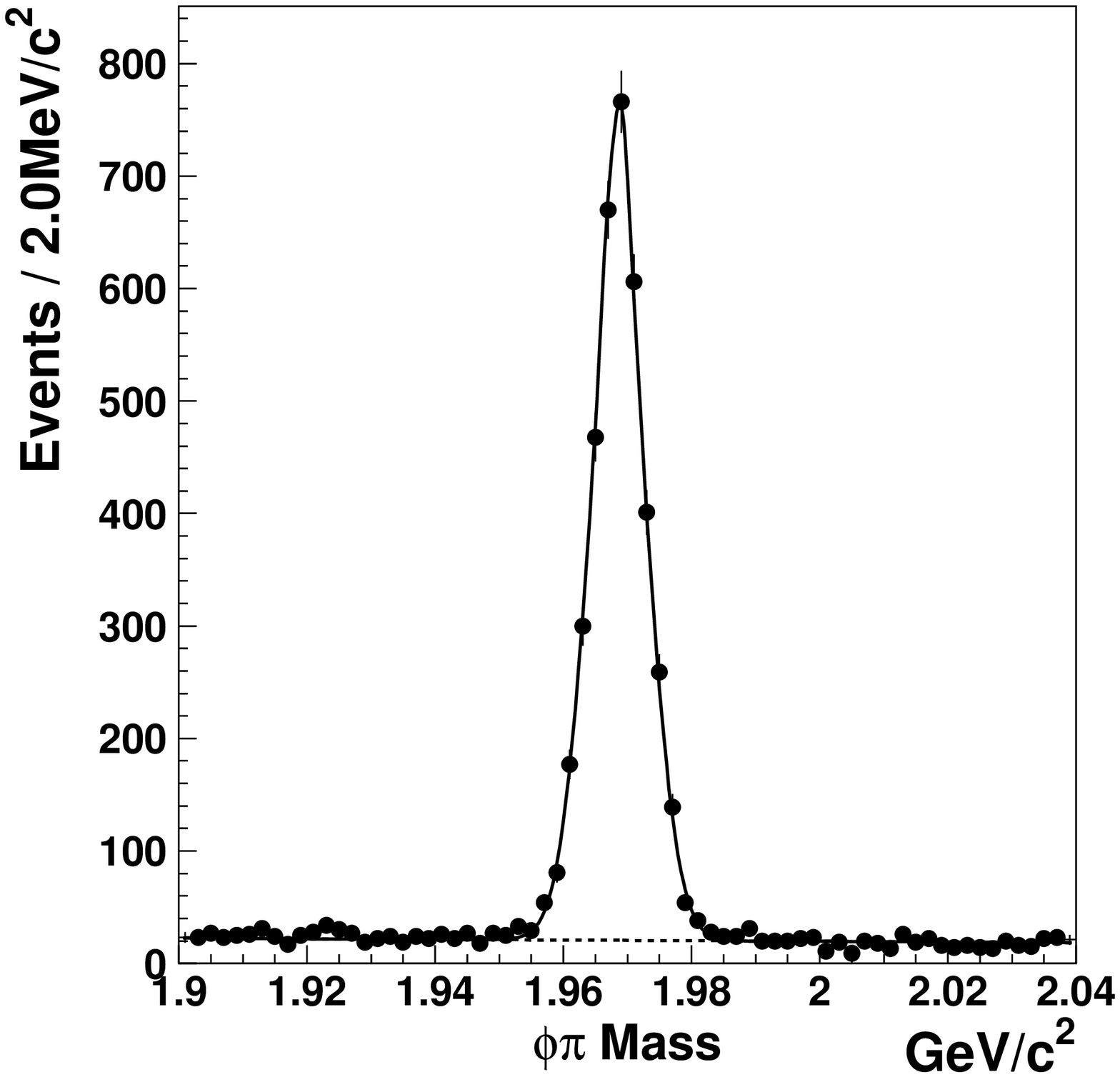}}
\end{flushright}
(c)\hspace{0.1mm}\vspace{-6.5mm}
\begin{flushright}
\resizebox{0.45\textwidth}{0.25\textwidth}{\includegraphics{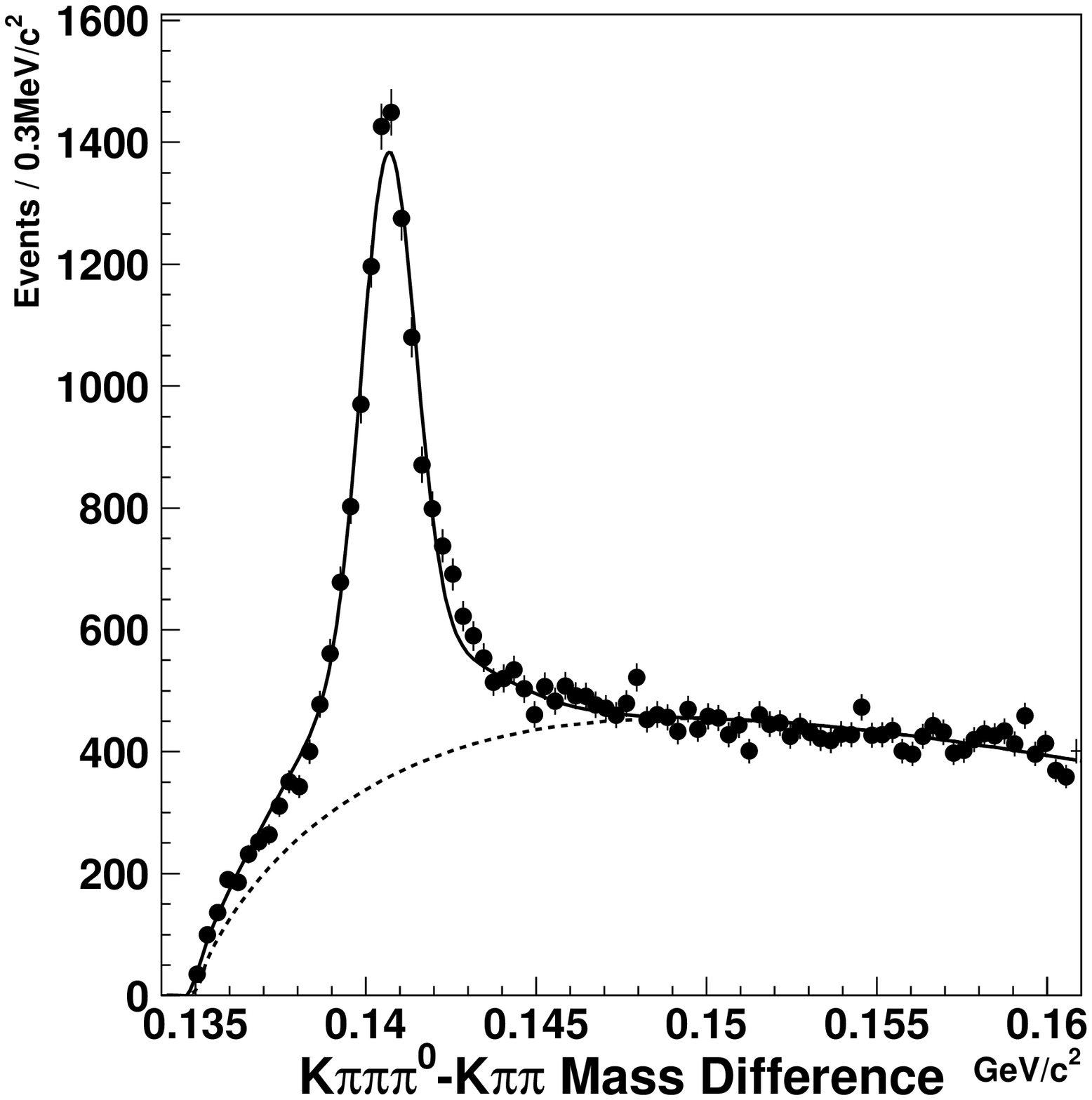}}
\end{flushright}
(d)\hspace{0.1mm}\vspace{-6.5mm}
\begin{flushright}
\resizebox{0.45\textwidth}{0.25\textwidth}{\includegraphics{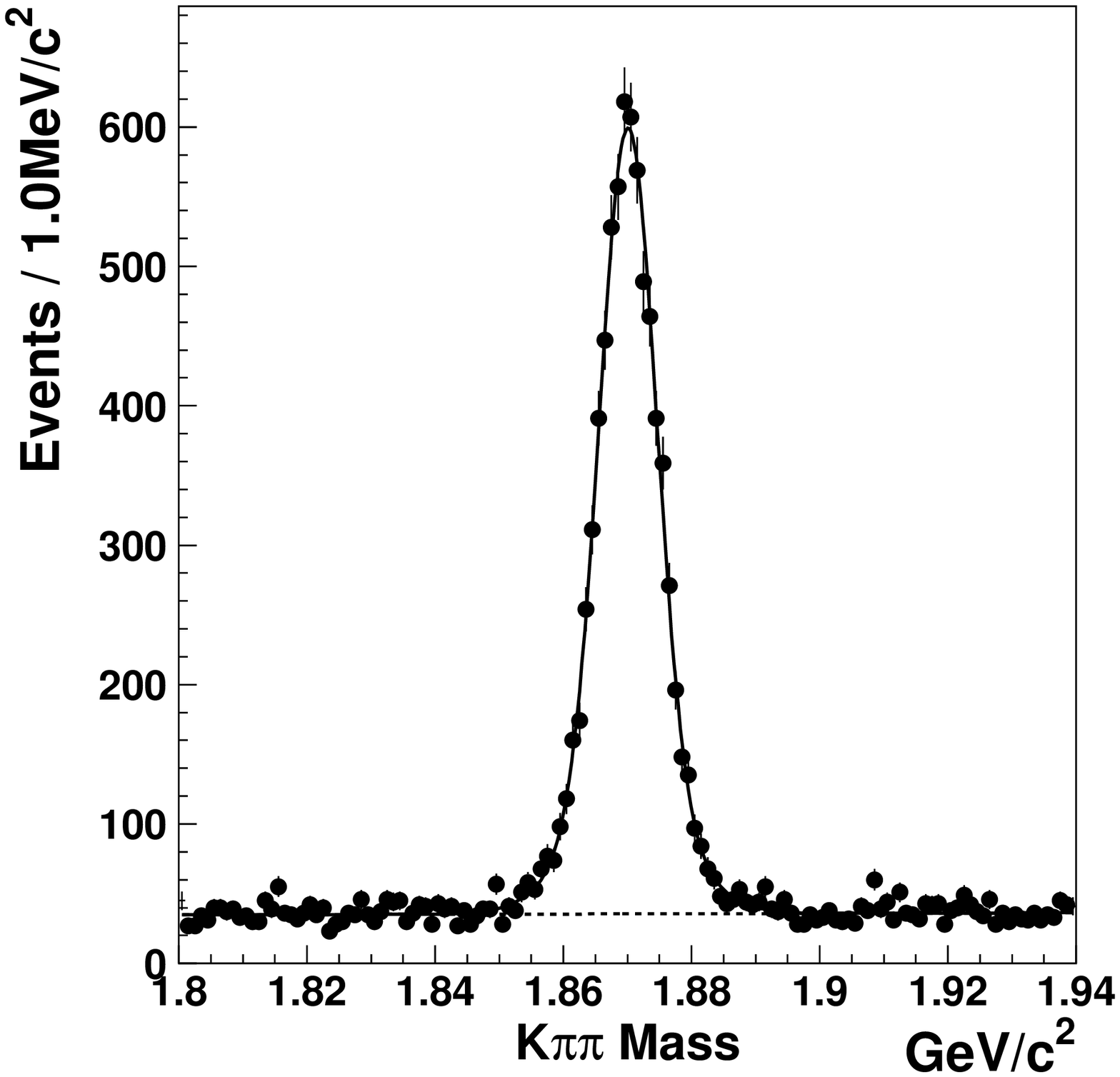}}
\end{flushright}
\caption{Mass and mass difference distributions:
(a) $D^0$($K^-\pi^+$), (b) $D_s^+$($\phi\pi^+$),
(c) $D^{*+}-D^+$ and (d) $D^+$($K^-\pi^+\pi^+$).}
\label{fig:mass}
\end{figure}

The charmed meson momentum in the $\Upsilon$(4S) rest frame is required to be greater than 2.5 \Gevc\
to eliminate $B\bar{B}$ events which do not come from the interaction region.
The other selection criteria, {\it e.g.} cuts on the decay angle, the helicity angle and
$\chi^2$/N.D.F of the track fit, are described in detail in the conference paper\cite{ICHEP2000} of ICHEP2000.
Figures~\ref{fig:mass} show mass and mass difference($m_{D^{*+}}-m_{D^+}$) distributions for
some decay chains of the $D^0$, $D^+$ and $D_s^+$.
We find $90601 \pm 387$ $\dzkpi$, $7451 \pm 118$ $\dzkk$,
$6953 \pm 99$ $\dpkpipi$ signals with $\dstarp\to\dplus\piz$ requirement, $1137\pm 35$ $\dpphipi$,
$3757\pm 54$ $\dsphipi$ and $2207\pm 68$ $\dskstk$ signals
within $3\sigma$ of the measured mean value.

The decay vertex($\vec2{x}_{dec}$) of the charmed meson is determined and then
the production vertex($\vec2{x}_{pro}$) is obtained by extrapolating the $D$
flight path to the interaction region of $e^+e^-$.
The projected decay length($L$) and the proper-time($t$) are obtained from
$L = (\vec2{x}_{pro}-\vec2{x}_{dec})\cdot\vec2{p}_D/|\vec2{p}_D|$ and
$t = Lm_D/c|\vec2{p}_D|$ respectively, where $\vec2{p}_D$ and $m_D$
are the momentum and mass of the charmed meson.

\section{Lifetime Fit}
An unbinned maximum likelihood fit is performed to extract the lifetimes.
The probability density function($P$) for each event consists of a signal term and
two background terms, representing components of the background
with non-zero lifetime and zero lifetime respectively.
The likelihood function($L$) is then given by
\begin{eqnarray*}
L &=& \prod_i P(t^i, \sigma_t^i, f_{SIG}^i) \\
&=& \prod_i [f_{SIG}^i\int^\infty_0 dt^\prime\frac{1}{\tau_{SIG}}e^{\frac{-t^\prime}
{\tau_{SIG}}}
R_{SIG}(t^i-t^\prime,\sigma_t^i) \\
&+&
(1-f_{SIG}^i)\int^\infty_0 dt^\prime
\{f_{\tau_{BG}}\frac{1}{\tau_{BG}}e^{\frac{-t^\prime}{\tau_{BG}}} \\
&+&(1-f_{\tau_{BG}})\delta(t^\prime)\}
R_{BG}(t^i-t^\prime,\sigma_t^i)],
\end{eqnarray*}
where $f_{SIG}^i$ and $f_{\tau_{BG}}$ are fractions for the signal and
the background with lifetime,
$\tau_{SIG}$ and $\tau_{BG}$ are the signal and background lifetimes,
$R_{SIG}$ and $R_{BG}$ are the resolution functions for the signal and the background,
and $t^i$, $\sigma_t^i$ are the measured proper-time, and its error, for each event.
The fraction $f_{SIG}^i$ is obtained based on the charmed meson mass for each event.
Since $f_{SIG}^i$ distributes an event to signal and background contributions properly,
we fit lifetimes in a wide range($\pm40$\Mevcsq) of the $D$ mass, that is, in signal and
sideband regions simultaneously.
The resolution functions $R_{SIG}$ and $R_{BG}$ are represented using
\begin{eqnarray*}
R(t,\sigma_t) &=& (1-f_{tail})\frac{1}{\sqrt{2\pi}S\sigma_t}
e^{-\frac{t^2}{2S^2\sigma_t^2}}\\
&+&
f_{tail}\frac{1}{\sqrt{2\pi}S_{tail}\sigma_t}
e^{-\frac{t^2}{2S_{tail}^2\sigma_t^2}},
\end{eqnarray*}
where $S$ and $S_{tail}$ are global scaling factors for the estimated error
$\sigma_t$ for the main and tail Gaussian distributions and
$f_{tail}$ is the fraction of the tail part.

\begin{figure}[H]
(a)\hspace{0.1mm}\vspace{-6.5mm}
\begin{flushright}
\resizebox{0.45\textwidth}{0.25\textwidth}{\includegraphics{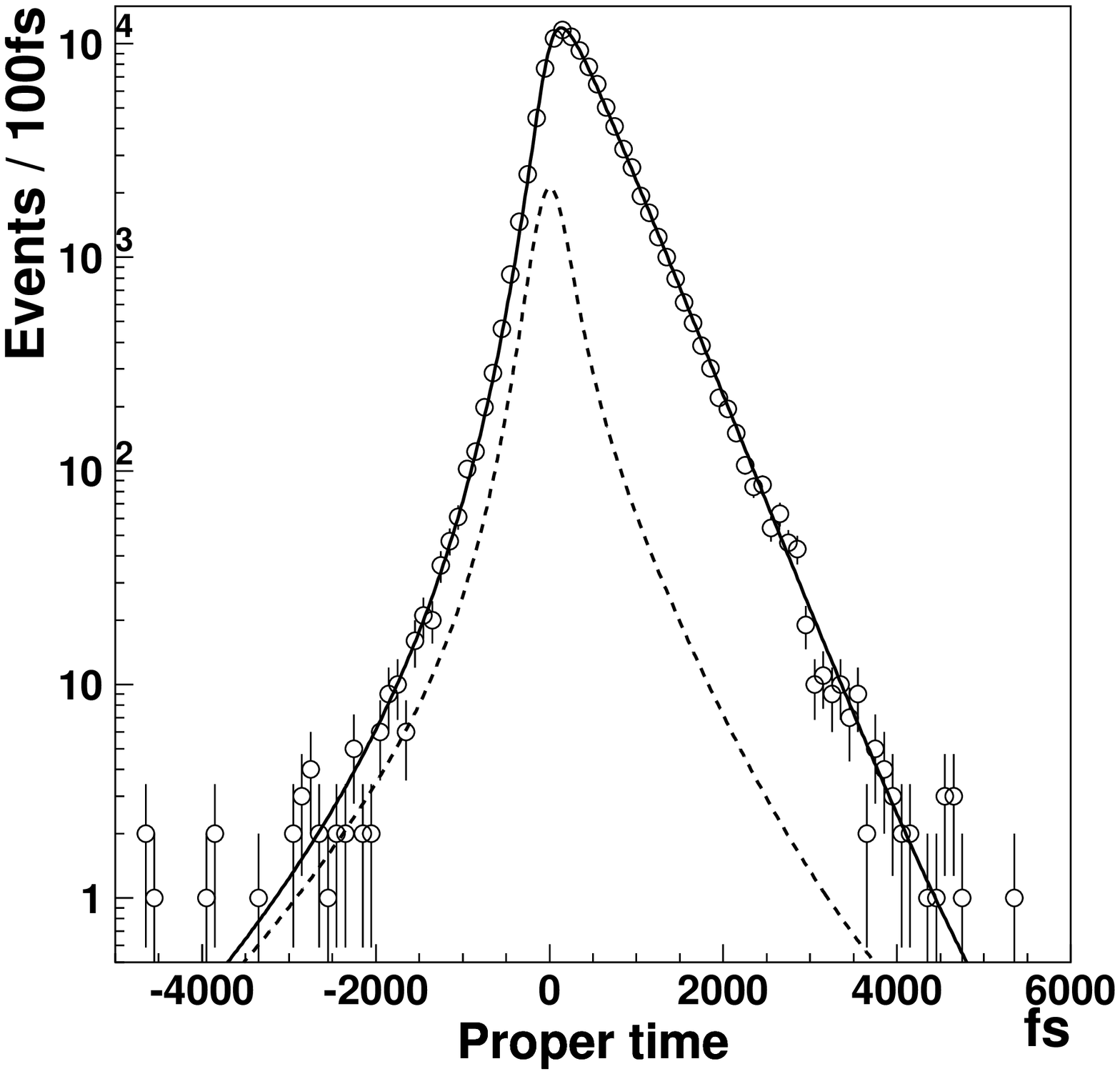}}
\end{flushright}
(b)\hspace{0.1mm}\vspace{-6.5mm}
\begin{flushright}
\resizebox{0.45\textwidth}{0.25\textwidth}{\includegraphics{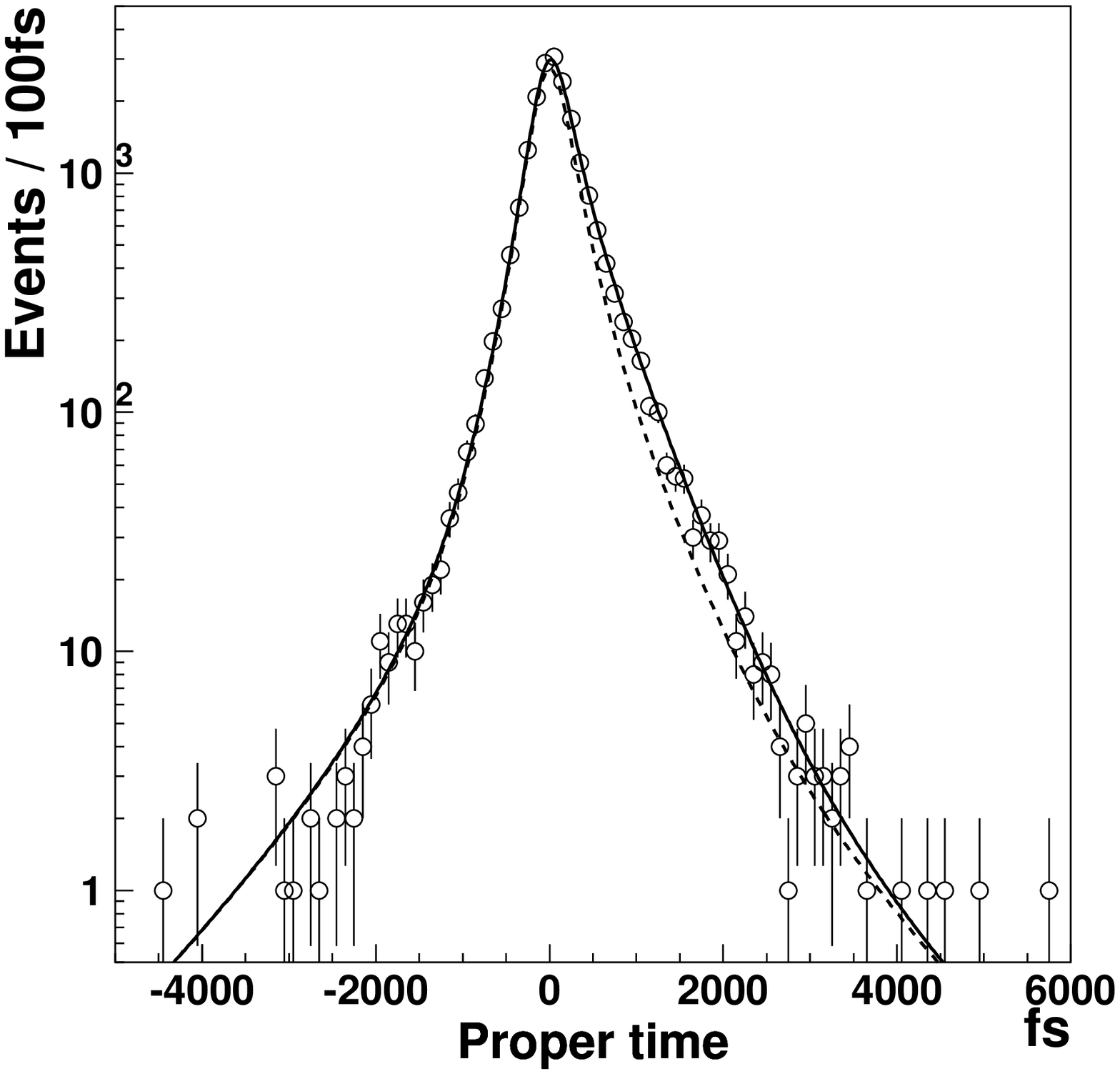}}
\end{flushright}
(c)\hspace{0.1mm}\vspace{-6.5mm}
\begin{flushright}
\resizebox{0.45\textwidth}{0.25\textwidth}{\includegraphics{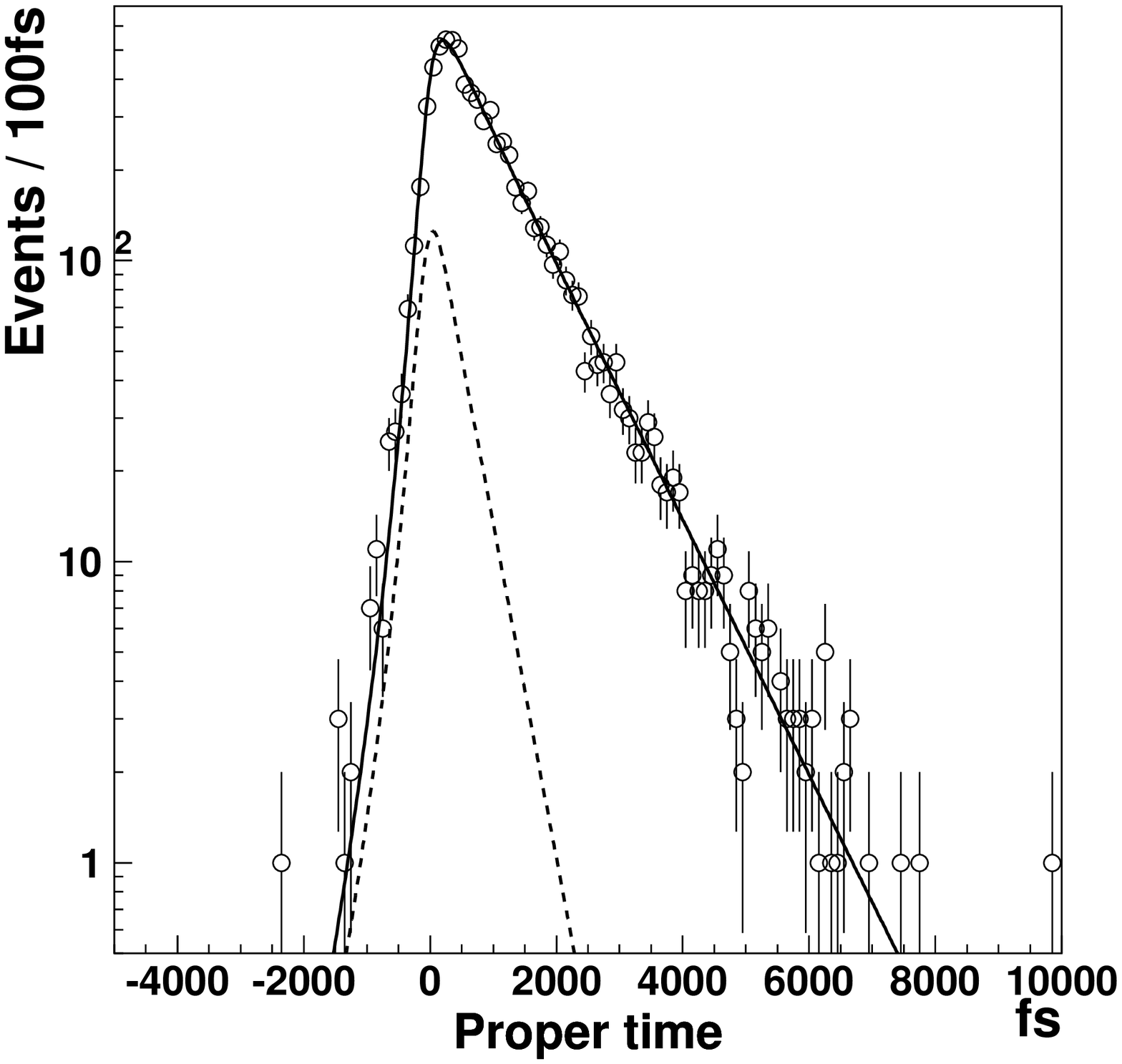}}
\end{flushright}
(d)\hspace{0.1mm}\vspace{-6.5mm}
\begin{flushright}
\resizebox{0.45\textwidth}{0.25\textwidth}{\includegraphics{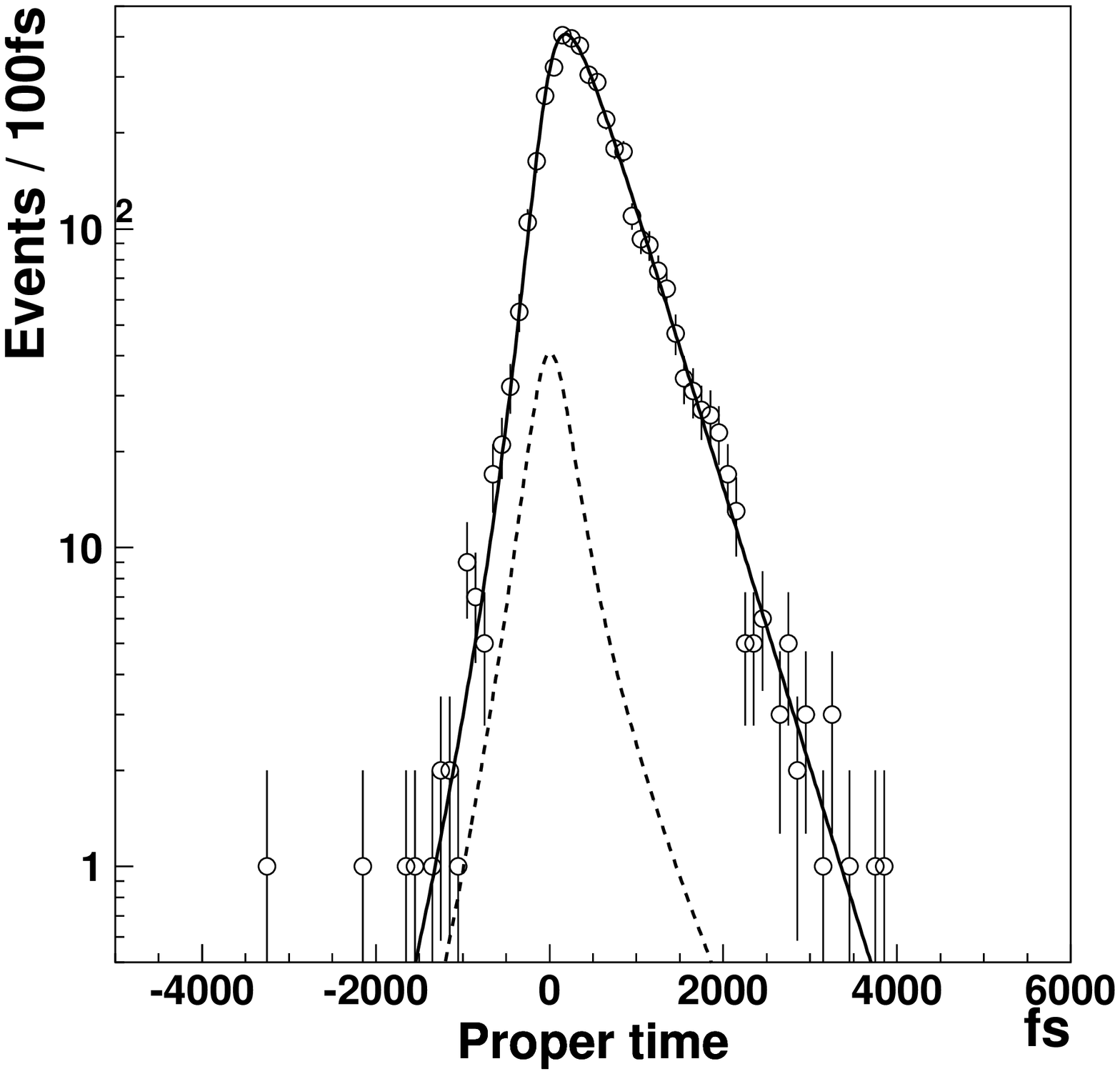}}
\end{flushright}
\caption{Proper-time distributions:
signal regions of $\dzkpi$(a),
$\dpkpipi$(c),
$\dsphipi$(d), and
a background region of $\dzkpi$(b),
where a circle is data, a solid curve is signal and background and
a dotted curve is background estimated from the fit result.}
\label{fig:lifetime_fit}
\end{figure}

Figures~\ref{fig:lifetime_fit} show the proper-time distributions and fit
results for $\dzkpi$, $\dpkpipi$ and $\dsphipi$.
We plot them in signal($<\pm3\sigma$) and sideband($>\pm3\sigma$) regions
in order to check whether our fits are good.
From the sideband region plot we can see that our background estimation is good.

\begin{figure}[H]
\begin{center}
\resizebox{0.40\textwidth}{0.30\textwidth}{\includegraphics{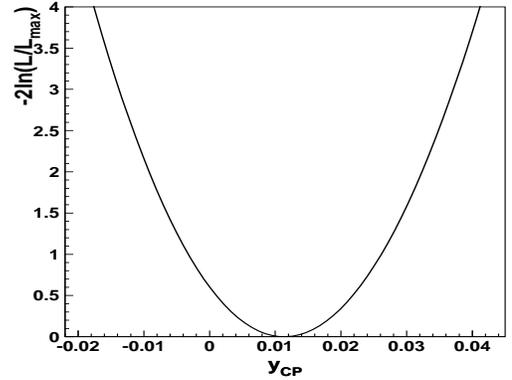}}
\caption{Log-likelihood as a function of $\ycp$.}
\label{fig:ycp_fit}
\end{center}
\end{figure}

We use a combined likelihood to obtain $\ycp$ and the $D^+$ and $D_s^+$ lifetimes,
since two final states are analyzed in each case.
This method makes it easier to estimate correlated systematic errors, {\it e.g.}
interaction point uncertainties.
The combined likelihoods are defined as
\begin{eqnarray*}
L_{\ycp}  &=& L_{\dzkpi}\cdot L_{\dzkk}, \\
L_{D^+}   &=& L_{\dpkpipi}\cdot L_{\dpphipi}, \\
L_{D_s^+} &=& L_{\dsphipi}\cdot L_{\dskstk}.
\end{eqnarray*}
In $L_{D^+}$ and $L_{D_s^+}$ a common $\tsig$ is used, \ie,
\begin{eqnarray*}
\tsig {}_{D^+}   &=& \tsig {}_{\dpkpipi} = \tsig {}_{\dpphipi} \\
\tsig {}_{D_s^+} &=& \tsig {}_{\dsphipi} = \tsig {}_{\dskstk}.
\end{eqnarray*}
In $L_{\ycp}$ the lifetime parameter of $\dzkk$, $\tsig {}_{\dzkk}$,
is calculated from 
\begin{eqnarray*}
\tsig {}_{\dzkk} &=& \tsig {}_{\dzkpi}/(1+\ycp).
\end{eqnarray*}
Figure~\ref{fig:ycp_fit} shows the log-likelihood, as a function of $\ycp$,
which is obtained from the combined fit method.

\section{Systematic Uncertainties}
We consider systematic uncertainties from the reconstruction of
the $D$ decay length and from the fit function.
We take into account
decay vertex uncertainties,
vertexing cut dependence,
uncertainties of the Monte Carlo correction,
the size and position uncertainties of the interaction point,
mass dependence of the proper-time,
$D$-mass sideband region dependence and so on.
The first, second and third items are the major sources of systematic error.
Since we are still studying these systematic uncertainties,
we do not quote them here.

\begin{table*}[t]
\caption{Comparison of our results with PDG world averages and recent measurements.}
\begin{center}
\begin{tabular}{ccccc}
\hline \hline
Experiment & $\tdz$~fs & $\tdp$~fs & $\tds$~fs & $\ycp$ \% \\ \hline
PDG\cite{PDG2000} & $412.6\pm2.8$ & $1051\pm13$ & $496^{+10}_{-9}$ &-- \\
E791    & $(413\pm3\pm4)^\dagger$ & -- & $(518\pm14\pm7)^\dagger$ & $0.8\pm2.9\pm1.0$ \\
CLEO    & $(408.5\pm4.1^{+3.5}_{-3.4})^\dagger$ & $(1034\pm22^{+10}_{-13})^\dagger$ & $(486\pm15\pm5)^\dagger$ & $-1.1\pm2.5\pm1.4$ \\
FOCUS   & $409.2\pm1.3^\ddagger$ & -- & $506\pm8^\ddagger$ & $3.42\pm1.39\pm0.74$ \\
Belle   & $\tdzresultns$ & $\tdpresultns$ & $\tdsresultns$ & $\yresultns$ \\ \hline \hline
\multicolumn{5}{l}{${}^\dagger$This result is included in the PDG2000 world average.}\\
\multicolumn{5}{l}{${}^\ddagger$No systematic error is given.}\\
\end{tabular}
\end{center}
\label{table:comparison}
\end{table*}

\section{Doubly Cabibbo-Suppressed Decay}
\begin{figure}[H]
\begin{center}
\resizebox{0.40\textwidth}{0.3\textwidth}{\includegraphics{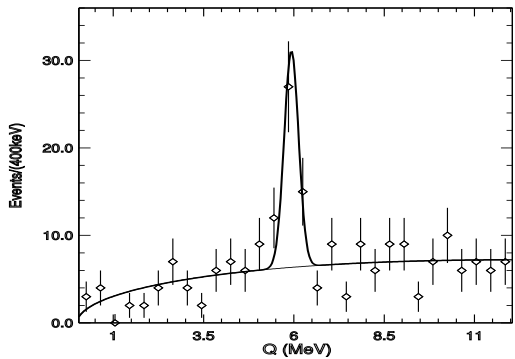}}
\caption{$Q$ distribution of wrong-sign decay $D^0\to K^+\pi^-$.}
\label{fig:ws}
\end{center}
\end{figure}

We are also studying wrong-sign decays $D^0\to K^+\pi^-$
to measure $x'$ and $y'$ from the proper-time distribution of wrong-sign decays.
We can see a clear peak in the $Q$($=m_{D^*}-m_{D^0}-m_{\pi}$) distribution for wrong-sign decays,
as shown in Figure~\ref{fig:ws} and are still studying
the various background components, and their distributions in ($Q$, $m_{D^0}$).

\section{Conclusions}
We have presented new measurements of charmed meson lifetimes using \intl~\Fb\ of
data sample collected with the Belle detector near the \UPS\ energy.
Unbinned maximum likelihood fits to proper-time distributions of fully reconstructed
charmed meson candidates yield results for the lifetime and the mixing parameter $\ycp$
as shown in Table~\ref{table:comparison}.
The measured value of $\ycp$ is consistent with zero.
We need more data to measure it more precisely and test the FOCUS result.
The statistical uncertainties on the lifetimes are better than those of the best
published measurements.
All results are preliminary.

\section*{Acknowledgments}
We gratefully acknowledge the efforts of the KEKB group in providing
us with excellent luminosity and running conditions and the
help with our computing and network systems provided by members
of the KEK computing research center.
We thank the staffs of KEK and collaborating institutions for their
contributions to this work,
and acknowledge support from the Ministry of Education, Science, Sports and Culture of Japan and
the Japan Society for the Promotion of Science.

\end{document}